\documentclass[showpacs,preprintnumbers,amsmath,amssymb]{revtex4}

\usepackage{graphicx}
\usepackage{bm}
\usepackage{epstopdf}
\usepackage{float}
\usepackage{dcolumn}
\usepackage{bm}
\usepackage{mathrsfs}
\usepackage{amsmath}

\begin{document}

\title{The problem of embedded eigenvalues for the Dirac equation in the Schwarzschild black hole metric}
\author{D. Batic}
\email{dbatic@pi.ac.ae}
\affiliation{%
Department of Mathematics,\\  The Petroleum Institute, Abu Dhabi, UAE 
}
 
\author{M. Nowakowski}
\email{mnowakos@uniandes.edu.co}
\affiliation{
Departamento de Fisica,\\ Universidad de los Andes, Cra.1E
No.18A-10, Bogota, Colombia
}
 
\author{K. Morgan}
\email{kirk.morgan02@uwimona.edu.jm}
\affiliation{
Department of Mathematics,\\ University of the West Indies, Kingston 6, Jamaica
}

\date{\today}

\begin{abstract}
We use the Dirac equation in a fixed black hole background and
different independent techniques to demonstrate the absence
of fermionic bound states around a Schwarzschild black hole.
In particular, we show that no embedded eigenvalues exist
which have been claimed for the case when the energy is less than
the particle's mass. We explicitly prove that the claims regarding
the embedded eigenvalues can be traced back to an oversimplified
approximation in the calculation. We conclude that no bound states
exist regardless the value of the mass.

\end{abstract}

\pacs{Valid PACS appear here}
\maketitle
\section{Introduction}
In the last fifteen years the problem of whether or not bound states exist for the Dirac equation in the presence of a black hole attracted a lot of attention in the scientific community. The existing literature on this topic is characterized by disagreeing results. For instance, \cite{Zecca,Coat} showed that bound states exist for a Fermion immersed in the geometry of a Schwarzschild black hole. Their argument relies on an approximation of the radial system emerging from the Dirac equation after separation of variables and on the construction of approximated solutions at the event horizon and far away from the black hole. Both of them end up with an approximated spectrum resembling that of a hydrogen-like atom. On the other hand, \cite{Lasen} proved that such bound states cannot exist and only resonances are admitted. In their approach the problem is tackled numerically after writing the Dirac equation in the Schwarzschild black hole (BH) metric with respect to a gauge that is well-suited to numerical solution. Furthermore, it as been shown in \cite{Finster0} that the Dirac equation does not admit normalizable, time-periodic solutions in a non-extreme Reissner-Nordstr\"{o}m black hole geometry where the proof relies on the Heisenberg Uncertainty Principle and the particular form of the Dirac current. A non-existence theorem regarding bound states for the Dirac equation in a Kerr-Newman geometry was obtained by \cite{Finster} where the authors introduced certain matching conditions for the spinor field across the horizons which in turn gave rise to a weak solution of the Dirac equation in the physical region of the maximal analytic extension of the Kerr-Newman solution. More precisely, they exploited the conservation and positivity of the Dirac current to show that because of the matching conditions, the only way in which a bound state solution of the Dirac equation can be normalizable is that each term in the Fourier expansion of the spinor field in time and the angular variable around the axis of symmetry, be identically zero. The case of the extreme Kerr metric has been treated by \cite{S} where necessary and sufficient conditions for the existence of bound states have been derived. However, since such conditions are expressed in terms of a complicated set of equations and inequalities, the author did not further study the problem of whether or not there exist values of the particle energy satisfying the aforementioned conditions. Instead of dealing with the conditions derived by \cite{S}, \cite{Nowak} proved the non-existence of bound states for the Dirac equation in the extreme Kerr geometry by applying the so-called Index Theorem for Dirac systems presented in \cite{L}. Finally, the non-existence of bound states for the Dirac equation in the presence of a Kerr-Newman-de Sitter black hole was showed in \cite{Cacciatori} adopting an operator theory approach. In the present work we prove that no bound states occur for the Dirac equation in the Schwarzschild BH metric. Our method is different from those applied in the existing literature on this topic and in a certain sense much simpler since it boils down to the analysis of a four-term recurrence relation. Our paper is structured as follows.  In Section I we give a short introduction motivating the importance of our findings. In Section II we introduce the Dirac equation in the Schwarzschild BH metric and we review the different forms (tetrad dependent) obtained for the radial system emerging after the Chadrasekhar ansatz has been applied. At this point a comment is in order. The completeness of the Chandrasekhar ansatz has been proved for the first time in \cite{F2003} where an integral representation for the Dirac propagator in terms of the solutions of the radial and angular ODEs arising by means
of the Chandrasekhar method of separation of variables has been derived. \cite{F2003} inferred the completeness of this ansatz from the aforementioned integral representation. An alternative proof of the completeness of the Chandrasekhar ansatz can also be found in \cite{BS}. In Section III rather than approximating the radial system as in \cite{Zecca,Coat}, where bound states were found due to the oversimplification of the problem, we decouple it and show that each radial spinor satisfies a generalized Heun equation \cite{S0,S1,B}. Then, we require that the radial spinors exhibit exponential decay asymptotically away from the black hole. This will be always the case if $\omega^2<m_e^2$ where $\omega$ is the energy of the particle and $m_e$ the mass of the Fermion. The next step consists in establishing whether or not  it is possible to find representations of the radial wave functions consisting of an exponential decaying at space-like infinity multiplied by a polynomial function. The presence of a polynomial function will ensure integrability at the event horizon. Similarly to the study of bound states for the hydrogen atom we employ a power series representation whose coefficients turn out to satisfy a four-term recurrence relation instead of a two-term recurrence relation as it is the case for the hydrogen atom. At this point we derive necessary and sufficient conditions ensuring that our recurrence relation breaks down in the sense that the first $N$ coefficients of the power series expansion do not vanish and the subsequent coefficients are all zero. Finally, we show by contradiction that there exist no real values of the energy of the Fermion satisfying the aforementioned conditions and hence, no bound states can occur. However, if one allows for complex values of the energy, then our conditions can be satisfied simultaneously and resonances appear in agreement with the results in \cite{Lasen}. In Section IV we give an alternative proof relying on the application of the Index Theorem for Dirac systems (see \cite{L}).

\section{The Dirac equation in the Schwarzschild BH metric}
The Schwarzschild space-time represents a static black hole immersed in an asymptotically Minkowski space-time. In coordinates $(t,r,\vartheta,\varphi)$ with $r>0$, $0\leq\vartheta\leq\pi$, $0\leq\varphi<2\pi$ the Schwarzschild BH metric is given 
by \cite{Inverno}
\begin{equation}\label{SLE}
ds^{2}=\left(1-\frac{2M}{r}\right)dt^2-\left(1-\frac{2M}{r}\right)^{-1}dr^2-r^2(d\vartheta^2+\sin^2{\vartheta}d\varphi^2),
\end{equation}
where $M$ is the mass of the black hole. A fermion of mass $m_e$ and charge $e$ in a curved space-time is described by the Dirac equation ($\hbar=c=G=1$) \cite{Page,Ch}
\begin{equation}\label{unopage}
\nabla_{AA^{'}}P^A+i\mu_{*}\overline{Q}_{A^{'}}=0,\quad
\nabla_{AA^{'}}Q^A+i\mu_{*}\overline{P}_{A^{'}}=0,\quad\mu_{*}=\frac{m_e}{\sqrt{2}},  
\end{equation}
where $\nabla_{AA^{'}}$ denotes covariant differentiation, and $(P^A,Q^{A^{'}})$ with $A=0,1,$ and $A^{'}=0^{'},1^{'}$ are the two-component spinors representing the wave function. According to \cite{Ch}, at each point of the space-time we can associate to the spinor basis a null tetrad $(\bm{\ell},\mathbf{n},\mathbf{m},\overline{\mathbf{m}})$ obeying the normalization and orthogonality relations 
\begin{equation}\label{tetrad_relations}
\bm{\ell}\cdot\mathbf{n}=1,\quad 
\mathbf{m}\cdot\overline{\mathbf{m}}=-1,\quad
\bm{\ell}\cdot\mathbf{m}=\bm{\ell}\cdot\overline{\mathbf{m}}=\mathbf{n}\cdot\mathbf{m}=\mathbf{n}\cdot\overline{\mathbf{m}}=0
\end{equation}
and hence, the Dirac equation can be rewritten as \cite{Ch}
\begin{eqnarray}
(D+\epsilon-\rho)P^0+(\overline{\delta}+\pi-\alpha)P^1&=&i\mu_{*}\overline{Q^{1^{'}}},\\
(\delta+\beta-\tau)P^0+(\Delta+\mu-\gamma)P^1&=&-i\mu_{*}\overline{Q^{0^{'}}},\\
(D+\epsilon-\rho)Q^{0^{'}}+(\overline{\delta}+\pi-\alpha)Q^{1^{'}}&=&i\mu_{*}\overline{P^{1}},\\
(\delta+\beta-\tau)Q^{0^{'}}+(\Delta+\mu-\gamma)Q^{1^{'}}&=&-i\mu_{*}\overline{P^{0}},
\end{eqnarray}
where
\begin{equation}
D=\ell^i\partial_i,\quad
\Delta=n^i\partial_i,\quad
\delta=m^i\partial_i,\quad
\overline{\delta}=\overline{m^i}\partial_i,\quad i=1,\cdots,4
\end{equation}
and the spin coefficients are given by
\begin{eqnarray}
\kappa&=&\gamma_{(3)(1)(1)},\quad
\sigma=\gamma_{(3)(1)(3)},\quad
\lambda=\gamma_{(2)(4)(4)},\quad
\nu=\gamma_{(2)(4)(2)},\quad
\rho=\gamma_{(3)(1)(4)},\quad
\mu=\gamma_{(2)(4)(3)},\\
\tau&=&\gamma_{(3)(1)(2)},\quad
\pi=\gamma_{(2)(4)(1)},\quad
\epsilon=\frac{1}{2}\left[\gamma_{(2)(1)(1)}+\gamma_{(3)(4)(1)}\right],\quad
\gamma=\frac{1}{2}\left[\gamma_{(2)(1)(2)}+\gamma_{(3)(4)(2)}\right],\\
\alpha&=&\frac{1}{2}\left[\gamma_{(2)(1)(4)}+\gamma_{(3)(4)(4)}\right],\quad
\beta=\frac{1}{2}\left[\gamma_{(2)(1)(3)}+\gamma_{(3)(4)(3)}\right]
\end{eqnarray}
with $(a)$ denoting the tetrad index and $a=1,\cdots,4$. The Ricci rotation coefficients are expressed as
\begin{equation}
\gamma_{(a)(b)(c)}=\frac{1}{2}\left[\lambda_{(a)(b)(c)}+\lambda_{(c)(a)(b)}-\lambda_{(b)(c)(a)}\right],\quad
\lambda_{(a)(b)(c)}=\left[e_{(b)i,j}-e_{(b)j,i}\right]e_{(a)}{}^ie_{(c)}{}^j,
\end{equation}
where
\begin{equation}
(e_{(a)}{}^i)=\left(
\begin{array}{cccc}
\ell^1 & \ell^2 & \ell^3 & \ell^4\\
n^1    & n^2    & n^3    & n^4 \\
m^1    & m^2    & m^3    & m^4 \\
\overline{m^1}    & \overline{m^2}    & \overline{m^3}    & \overline{m^4}
\end{array}
\right)
\end{equation}
and its inverse is denoted by $(e^{(b)}{}_i)$. We recall that tetrad indices label rows and tensor indices label columns. Taking into account that $e_{(a)i}=g_{ik}e_{(a)}{}^k$ and assuming a diagonal metric tensor $g$ we find that
\begin{equation}
(e_{(a)}i)=\left(
\begin{array}{cccc}
\ell_1 & \ell_2 & \ell_3 & \ell_4\\
n_1    & n_2    & n_3    & n_4 \\
m_1    & m_2    & m_3    & m_4 \\
\overline{m_1}    & \overline{m_2}    & \overline{m_3}    & \overline{m_4}
\end{array}
\right)
\end{equation}
We have different null tetrads leading to different expressions of the spin coefficients. For instance, \cite{Ch,Zecca0} use the so-called Kinnersley tetrad
\begin{equation}\label{K}
\bm{\ell}=\bm{e}_{(1)}=\left(\frac{r^2}{\Delta_r},1,0,0\right),\quad
\bm{n}=\bm{e}_{(2)}=\left(\frac{1}{2},-\frac{\Delta_r}{2r^2},0,0\right),\quad
\bm{m}=\bm{e}_{(3)}=\left(0,0,\frac{1}{r\sqrt{2}},\frac{i}{r\sqrt{2}\sin{\vartheta}}\right)
\end{equation}
with $\Delta_r=r^2-2Mr$. 
\begin{table}[ht]\label{tt}
\begin{center}
  \begin{tabular}{ | c | c | c | c | c | c | c | c | c | c | c | c | c |}
    \hline
Vierbein                        & $\kappa$ & $\sigma$ & $\lambda$ & $\nu$ & $\rho$ & $\mu$ & $\tau$ & $\pi$ & $\epsilon$ & $\gamma$ & $\beta$ & $\alpha$\\ \hline
  Kinnersley (\ref{K})          &  $0$     & $0$      & $0$       & $0$   & $-\frac{1}{r}$     & $\frac{2M-r}{2r^2}$     & $0$   &$0$    & $0$    &  $\frac{M}{2r^2}$        & $\frac{\cot{\vartheta}}{2\sqrt{2}r}$ & $-\beta$  \\ \hline
  Carter (\ref{tetrad})         &  $0$     & $0$      & $0$       & $0$   & $\frac{\sqrt{\Delta_r}}{r^2\sqrt{2}}$       & $\rho$      &$0$        & $0$    &   $-\frac{M}{2\sqrt{2}r\sqrt{\Delta_r}}$         & $\epsilon$  & $\frac{\cot{\vartheta}}{2\sqrt{2}r}$ & $-\beta$  \\
    \hline
  \end{tabular}
\end{center}
\caption{Table of spin coefficients for different choices of the null tetrad}
\end{table}
Furthermore, one can also use the Carter tetrad \cite{Cart1,Cart} which allows for a more elegant treatment of the separation problem and leads to a simpler form of the radial system than those derived in \cite{Ch,Zecca0}. To this purpose we first introduce a vierbein $\{\widetilde{\bm{e}}_{(a)}\}$ with $a=1,\cdots,4$ such that $g=\eta_{(a)(b)}\widetilde{\bm{e}}^{(a)}\otimes\widetilde{\bm{e}}^{(b)}$ with $\eta_{(a)(b)}=\mbox{diag}(1,-1,-1,-1)$. For the Schwarzschild BH metric we have
\begin{equation}
\widetilde{\bm{e}}^{(1)}=\left(\frac{\sqrt{\Delta_r}}{r},0,0,0\right),\quad
\widetilde{\bm{e}}^{(2)}=\left(0,\frac{r}{\sqrt{\Delta_r}},0,0\right),\quad
\widetilde{\bm{e}}^{(3)}=(0,0,r,0),\quad
\widetilde{\bm{e}}^{(4)}=(0,0,0,-r\sin{\vartheta}).
\end{equation}
With the help of $(5.119)$ in \cite{Carter} we can construct the following symmetric null tetrad \cite{BS,Davis}
\begin{equation}\label{tetrad}
\bm{\ell}=\left(\frac{r}{\sqrt{2\Delta_r}},-\frac{1}{r}\sqrt{\frac{\Delta_r}{2}},0,0\right),\quad
\bm{n}=\left(\frac{r}{\sqrt{2\Delta_r}},\frac{1}{r}\sqrt{\frac{\Delta_r}{2}},0,0\right),\quad
\bm{m}=\left(0,0,\frac{1}{r\sqrt{2}},\frac{i}{r\sqrt{2}\sin{\vartheta}}\right).
\end{equation}
The corresponding spin coefficients are listed in Table~I. We complete this section by giving an overview of the different but equivalent radial systems emerging from the Dirac equation after separation of variables when the Kinnersley and Carter tetrads are used. First of all, \cite{Ch} separates the Dirac equation in the Schwarzschild BH metric by choosing the Kinnersley tetrad and by making the ansatz
\begin{eqnarray}
P^0&=&\frac{e^{i(\omega t+k\varphi)}}{r\sqrt{2}}R_{-}(r)S_{-}(\vartheta),\quad
P^1=e^{i(\omega t+k\varphi)}R_{+}(r)S_{+}(\vartheta),\\
Q^{0^{'}}&=&-\frac{e^{-i(\omega t+k\varphi)}}{r\sqrt{2}}\overline{R_{-}(r)}~\overline{S_{+}(\vartheta)},\quad
Q^{1^{'}}=e^{-i(\omega t+k\varphi)}\overline{R_{+}(r)}~\overline{S_{-}(\vartheta)},
\end{eqnarray}
where $\omega$ is the energy of the particle and $k=\pm 1/2,\pm 3/2,\cdots$ is the azimuthal quantum number \cite{BS}. This procedure leads to the following radial and angular systems
\begin{equation}
\left(
\begin{array}{cc}
D_0 & -(\lambda+im_e r)\\
-(\lambda-im_e r) &\sqrt{\Delta_r}D^\dagger_0\sqrt{\Delta_r}
\end{array}
\right)\left(
\begin{array}{c}
R_{-}\\
R_{+}
\end{array}
\right)=0,\quad
\left(
\begin{array}{cc}
\mathcal{L}^\dagger_{\frac{1}{2}} & -\lambda\\
\lambda &\mathcal{L}_{\frac{1}{2}}
\end{array}
\right)\left(
\begin{array}{c}
S_{-}\\
S_{+}
\end{array}
\right)=0,
\end{equation}
where
\begin{equation}
D_0=\frac{d}{dr}+i\frac{\omega r^2}{\Delta_r},\quad D_0^\dagger=\frac{d}{dr}-i\frac{\omega r^2}{\Delta_r},\quad
\mathcal{L}_{\frac{1}{2}}=\frac{d}{d\vartheta}+\frac{k}{\sin{\vartheta}}+\frac{1}{2}\cot{\vartheta},\quad
\mathcal{L}^\dagger_{\frac{1}{2}}=\frac{d}{d\vartheta}-\frac{k}{\sin{\vartheta}}+\frac{1}{2}\cot{\vartheta}.
\end{equation}
The spectrum of the angular problem is purely discrete and the eigenvalues are given as follows \cite{Bat2005}
\begin{equation}
\lambda_j(k)=\mbox{sgn}(j)\left(|k|-\frac{1}{2}+|j|\right),\quad j\in\mathbb{Z}\backslash\{0\}.
\end{equation}
Finally, the radial system can be rewritten as 
\begin{eqnarray}
\frac{dR_{-}}{dr}+i\frac{\omega r}{r-2M}R_{-}&=&(\lambda+i m_e r)R_{+},\\
\frac{dR_{+}}{dr}+\left(\frac{r-M}{r^2-2Mr}-i\frac{\omega r}{r-2M}\right)R_{+}&=&\frac{\lambda-im_e r}{r^2-2Mr}R_{-}.
\end{eqnarray}
The Dirac equation in the Schwarzschild BH metric has been also separated in \cite{Zecca0} where the Kinnersley tetrad has been adopted but the initial ansatz for the spinor is somewhat different from that made in \cite{Ch}. In this case we let
\begin{eqnarray}
P^0&=&\frac{e^{i(\omega t+k\varphi)}}{r}H_{1}(r)S_{1}(\vartheta),\quad
P^1=\frac{e^{i(\omega t+k\varphi)}}{r}H_{2}(r)S_{2}(\vartheta),\\
Q^{0^{'}}&=&-\frac{e^{-i(\omega t+k\varphi)}}{r}\overline{H_{1}(r)}~\overline{S_{2}(\vartheta)},\quad
Q^{1^{'}}=\frac{e^{-i(\omega t+k\varphi)}}{r}\overline{H_{2}(r)}~\overline{S_{1}(\vartheta)}.
\end{eqnarray}
and after substitution into the Dirac equation we obtain the following radial and angular systems
\begin{equation}
\left(
\begin{array}{cc}
r\sqrt{2}D_0 & \lambda-im_e r\\
\lambda+im_e r &\frac{\Delta_r}{r\sqrt{2}}\left(D^\dagger_0+\frac{M}{\Delta_r}\right)
\end{array}
\right)\left(
\begin{array}{c}
H_{1}\\
H_{2}
\end{array}
\right)=0,\quad
\left(
\begin{array}{cc}
\mathcal{L}^{+} & \lambda\\
-\lambda &\mathcal{L}^{-}
\end{array}
\right)\left(
\begin{array}{c}
S_{1}\\
S_{2}
\end{array}
\right)=0,
\end{equation}
where $D^0$ and $D^\dagger_0$ are defined as in \cite{Ch}. Moreover, $\mathcal{L}^{+}=\mathcal{L}^\dagger_{\frac{1}{2}}$ and $\mathcal{L}^{-}=\mathcal{L}_{\frac{1}{2}}$. Finally, the radial system can be rewritten as 
\begin{eqnarray}
\frac{dH_{1}}{dr}+i\frac{\omega r}{r-2M}H_{1}&=&\left(i\mu_{*}-\frac{\lambda}{r\sqrt{2}}\right)H_{2},\\
\frac{dH_{2}}{dr}+\left(\frac{M}{r^2-2Mr}-i\frac{\omega r}{r-2M}\right)H_{2}&=&-\frac{2r}{r-2M}\left(i\mu_{*}+\frac{\lambda}{r\sqrt{2}}\right)H_{1}.
\end{eqnarray}
We end this section with a short derivation of the radial system when the Carter tetrad is chosen. The corresponding spin coefficients are given in Table~I. Let
\begin{eqnarray}
P^0&=&\frac{e^{i(\omega t+k\varphi)}}{\sqrt{r}\sqrt[4]{\Delta_r}}R_{+}(r)S_{+}(\vartheta),\quad
P^1=\frac{e^{i(\omega t+k\varphi)}}{\sqrt{r}\sqrt[4]{\Delta_r}}R_{-}(r)S_{-}(\vartheta),\\
Q^{0^{'}}&=&-\frac{e^{-i(\omega t+k\varphi)}}{\sqrt{r}\sqrt[4]{\Delta_r}}\overline{R_{-}(r)}~\overline{S_{+}(\vartheta)},\quad
Q^{1^{'}}=\frac{e^{-i(\omega t+k\varphi)}}{\sqrt{r}\sqrt[4]{\Delta_r}}\overline{R_{+}(r)}~\overline{S_{-}(\vartheta)}.
\end{eqnarray}
Then, the Dirac equation splits into the following radial and angular systems
\begin{equation}\label{CRAS}
\left(
\begin{array}{cc}
\sqrt{\Delta_r}D_0 & -(\lambda+im_e r)\\
-(\lambda-im_e r) &\sqrt{\Delta_r}D^\dagger_0
\end{array}
\right)\left(
\begin{array}{c}
R_{-}\\
R_{+}
\end{array}
\right)=0,\quad
\left(
\begin{array}{cc}
\mathcal{L}^{\frac{1}{2}} & -\lambda\\
\lambda &\mathcal{L}^{\dagger}_{\frac{1}{2}}
\end{array}
\right)\left(
\begin{array}{c}
S_{-}\\
S_{+}
\end{array}
\right)=0,
\end{equation}
where $D^0$, $D^\dagger_0$, $\mathcal{L}^{\frac{1}{2}}$, and $\mathcal{L}^{\dagger}_{\frac{1}{2}}$ are defined as in \cite{Ch}. The following property of the radial spinors
\begin{equation}\label{property}
\overline{R_{-}(r)}=R_{+}(r),\quad
\overline{R_{+}(r)}=R_{-}(r)
\end{equation}
becomes evident if we rewrite the radial system as
\begin{eqnarray}
\frac{dR_{-}}{dr}+i\frac{\omega r}{r-2M}R_{-}&=&\frac{\lambda+im_e r}{\sqrt{r^2-2Mr}}R_{+},\label{r1}\\
\frac{dR_{+}}{dr}-i\frac{\omega r}{r-2M}R_{+}&=&\frac{\lambda-im_e r}{\sqrt{r^2-2Mr}}R_{-}.\label{r2}
\end{eqnarray} 
Note that property (\ref{property}) could not have become transparent if we would have adopted the Kinnersley tetrad. This also implies that the solution of the Dirac equation will read
\begin{eqnarray}
P^0&=&\frac{e^{i(\omega t+k\varphi)}}{\sqrt{r}\sqrt[4]{\Delta_r}}R_{+}(r)S_{+}(\vartheta),\quad
P^1=\frac{e^{i(\omega t+k\varphi)}}{\sqrt{r}\sqrt[4]{\Delta_r}}R_{-}(r)S_{-}(\vartheta),\\
Q^{0^{'}}&=&-\frac{e^{-i(\omega t+k\varphi)}}{\sqrt{r}\sqrt[4]{\Delta_r}}R_{+}(r)~\overline{S_{+}(\vartheta)},\quad
Q^{1^{'}}=\frac{e^{-i(\omega t+k\varphi)}}{\sqrt{r}\sqrt[4]{\Delta_r}}R_{-}(r)~\overline{S_{-}(\vartheta)}.
\end{eqnarray}
If we let
\begin{equation}
P^0=\frac{F_1}{\sqrt{r}\sqrt[4]{\Delta_r}},\quad
P^1=\frac{F_2}{\sqrt{r}\sqrt[4]{\Delta_r}},\quad
Q^{0^{'}}=-\frac{\overline{G_2}}{\sqrt{r}\sqrt[4]{\Delta_r}},\quad
Q^{1^{'}}=\frac{\overline{G_1}}{\sqrt{r}\sqrt[4]{\Delta_r}},
\end{equation}
the Dirac equation in the presence of a Schwarzschild manifold can be written as
\begin{equation}\label{I}
(\mathcal{R}+\mathcal{A})\Psi=0,\quad \Psi=(F_1,F_2,G_1,G_2)^T
\end{equation}
with
\begin{equation}
\mathcal{R}=\left(
\begin{array}{cccc}
-im_e r & 0 & \sqrt{\Delta_r}\mathcal{D}_+ & 0\\
0 & -im_e r & 0 &-\sqrt{\Delta_r}\mathcal{D}_- \\
-\sqrt{\Delta_r}\mathcal{D}_- & 0 & -im_e r & 0\\
0 & \sqrt{\Delta_r}\mathcal{D}_+ & 0 & -im_e r
\end{array}
\right),\quad
\mathcal{A}=\left(
\begin{array}{cccc}
0 & 0 & 0 & -\mathcal{L}_-\\
0 & 0 & -\mathcal{L}_+ & 0\\
0 & \mathcal{L}_- & 0 & 0\\
\mathcal{L}_+ & 0 & 0 & 0
\end{array}
\right)
\end{equation}
and differential operators
\begin{equation}
\mathcal{D}_{\pm}=\partial_r\pm\frac{r^2}{\Delta_r}\partial_t,\quad
\mathcal{L}_{\pm}=\partial_\vartheta\pm\frac{i}{\sin{\vartheta}}\partial_\varphi+\frac{1}{2}\cot{\vartheta}.
\end{equation}
From (\ref{I}) we immediately obtain the Dirac equation in Hamiltonian form, namely
\begin{equation}\label{DE}
i\partial_t\Psi=H\Psi,\quad H=i\frac{\Delta_r}{r^2}\mbox{diag}(-\partial_r,\partial_r,\partial_r,-\partial_r)-\frac{m_e}{r}\sqrt{\Delta_r}A+i\frac{\sqrt{\Delta_r}}{r^2}B
\end{equation}
with
\begin{equation}
A=\left(
\begin{array}{cccc}
0&0&1&0\\
0&0&0&1\\
1&0&0&0\\
0&1&0&0
\end{array}
\right),\quad
B=\left(
\begin{array}{cccc}
0 & \mathcal{L}_- & 0 & 0\\
\mathcal{L}_+ & 0 & 0 & 0\\
0 & 0 & 0 & -\mathcal{L}_-\\
0 & 0 & -\mathcal{L}_+ & 0
\end{array}
\right).
\end{equation}
According to \cite{S} on the region $r>2M$ and on the hypersurfaces $t=\mbox{const}$ we can introduce the following inner product 
\begin{equation}\label{stern}
\langle\Psi,\Phi\rangle=\int_{2M}^{+\infty} dr~\frac{r^2}{\Delta_r}\int_0^\pi d\vartheta~\sin{\vartheta}\int_0^{2\pi}d\varphi{\Psi^{*}}^T(t,r,\vartheta,\varphi)\Phi(t,r,\vartheta,\varphi).
\end{equation}
It is not difficult to verify that the Hamiltonian $H$ is symmetric or formally self-adjoint with respect to the inner product (\ref{stern}). This in turn will imply that $\omega\in\mathbb{R}$. In what follows we say that a time-periodic solution
\begin{equation}
\Psi(t,r,\vartheta,\varphi)=e^{-i\omega t}\Psi_0(r,\vartheta,\varphi),\quad\omega\in\mathbb{R},\quad\Psi_0\neq 0
\end{equation}
of the Dirac equation (\ref{DE}) is a bound state if it is normalizable, that is
\begin{equation}
\langle\Psi,\Psi\rangle=\langle\Psi_0,\Psi_0\rangle<\infty,
\end{equation}
where $\langle\cdot,\cdot\rangle$ is given by (\ref{stern}). If such a solution exists, we say that $\omega$ is an eigenvalue of the Hamiltonian $H$ for the eigenspinor $\Psi_0$ and $\omega$ represents the energy of the bound state. Furthermore, if we introduce the Chandrasekhar ansatz
\begin{equation}\label{an}
\Psi_0(r,\vartheta,\varphi)=e^{ik\varphi}\left(
\begin{array}{c}
R_+(r)S_+(\vartheta)\\
R_-(r)S_-(\vartheta)\\
R_-(r)S_+(\vartheta)\\
R_-(r)S_-(\vartheta)
\end{array}
\right)
\end{equation}
with $k=\pm 1/2,\pm 3/2,\cdots$, we only need to investigate the radial and angular systems (\ref{CRAS}). The angular system has been thoroughly studied in \cite{Bat2005} where eigenvalues and associated eigenfunctions have been computed. Note that after separation of variables of the Dirac equation $\omega\in\mathbb{R}$ will be an energy eigenvalue of (\ref{DE}) if there exists some $\lambda\in\mathbb{R}$ and non trivial solutions
\begin{equation}
R(r)=\left(
\begin{array}{c}
R_-(r)\\
R_+(r)
\end{array}
\right),\quad
S(\vartheta)=\left(
\begin{array}{c}
S_-(\vartheta)\\
S_+(\vartheta)
\end{array}
\right),\quad r>2M,\quad\vartheta\in(0,\pi) 
\end{equation}
satisfying the normalization conditions
\begin{equation}\label{seven}
\int_{2M}^{+\infty} dr~\frac{r^2}{\Delta_r} |R(r)|^2<\infty,\quad
\int_0^\pi d\vartheta~\sin{\vartheta}~|S(\vartheta)|^2<\infty.
\end{equation}
By means of (\ref{an}) it is straightforward to verify that
\begin{equation}
\langle\Psi,\Psi\rangle=2\pi\left(\int_{2M}^{+\infty} dr~\frac{r^2}{\Delta_r} |R(r)|^2\right)\left(\int_0^\pi d\vartheta~\sin{\vartheta}~|S(\vartheta)|^2\right)<\infty,
\end{equation}
whenever (\ref{seven}) is satisfied and moreover, the approach followed here allows to reduce the eigenvalue equation $H\Psi_0=\omega\Psi_0$ and the normalization condition $\langle\Psi_0,\Psi_0\rangle$ to a pair of boundary value problems coupled by the separation constant $\lambda$. From \cite{Bat2005} we already know that the second inequality in (\ref{seven}) is satisfied for the eigenfunctions associated to the eigenvalues of the angular problem. Therefore, in the next two sections we will investigate whether or not there exist solutions of the radial problem such that the first inequality in (\ref{seven}) is satisfied.

\section{Non existence of Fermionic bound states in the Schwarzschild BH metric}
\cite{Zecca,Coat} claimed that bound states solutions for the Dirac equation in the Schwarzschild BH metric exist. The main idea behind their method is to study approximated solutions of the radial system close to the event horizon and asymptotically at space-like infinity. In particular, they obtained an asymptotic energy spectrum by requiring that the Kummer function describing the radial spinor asymptotically at infinity reduces to a polynomial function. The emergence of these bound states is a result of the oversimplification of the original system of differential equations governing the radial problem. This becomes particularly clear, if one studies the original radial system without introducing any sort of approximation. For this reason, we first reduce the system of differential equations satisfied by the radial spinors to a couple of generalized Heun equations  (see \cite{B,S0,S1}), despite the claims in \cite{Zecca0,Zecca} that such a radial system cannot be reduced to any known equation of mathematical physics, and then, we derive a set of necessary and sufficient conditions for the existence of bound state solutions for spin-$1/2$ particles in the Schwarzschild geometry. Finally, we prove that such a set of conditions can never be satisfied if the spectral parameter $\omega$ is real. If $\omega$ is allowed to be complex, polynomial solutions exist, but in this case instead of bound states, resonances will occur in agreement with the findings of \cite{Lasen}. Last but not least, our novel approach 
leads to a non-existence result in line with the findings of \cite{Cacciatori,Finster,Nowak} where the authors reached the same conclusion in the framework of operator theory. To this purpose, let $\Omega=2M\omega$ and $\mu=2Mm_e$. Furthermore, we introduce the rescaled radial variable $\rho=r/(2M)$ with $\rho\in(1,+\infty)$. If we set
\begin{equation}
f(\rho)=\frac{\rho}{\rho-1},\quad g(\rho)=\frac{\lambda+i\mu\rho}{\sqrt{\rho^2-\rho}},
\end{equation}
then, the radial system represented by (\ref{r1}) and (\ref{r2}) can be rewritten as follows
\begin{equation}
\frac{dR_{-}}{d\rho}+i\Omega f(\rho)R_{-}(\rho)=g(\rho)R_{+}(\rho),\quad
\frac{dR_{+}}{d\rho}-i\Omega f(\rho)R_{+}(\rho)=\overline{g(\rho)}R_{-}(\rho) 
\end{equation}
and it decouples into the following second order linear differential equations
\begin{eqnarray}
R^{''}_{-}(\rho)-\frac{g^{'}(\rho)}{g(\rho)}R^{'}_{-}(\rho)+\left[\Omega^2 f^2(\rho)-|g(\rho)|^2+i\Omega\left(f^{'}(\rho)-f(\rho)\frac{g^{'}(\rho)}{g(\rho)}\right)\right]R_{-}(\rho)&=&0,\\
R^{''}_{+}(\rho)-\frac{\overline{g^{'}(\rho)}}{\overline{g(\rho)}}R^{'}_{+}(\rho)+\left[\Omega^2 f^2(\rho)-|g(\rho)|^2-i\Omega\left(f^{'}(\rho)-f(\rho)\frac{\overline{g^{'}(\rho)}}{\overline{g(\rho)}}\right)\right]R_{+}(\rho)&=&0,
\end{eqnarray}
where a prime denotes differentiation with respect to the independent variable $\rho$. In order to bring the above differential equations into the form of a generalized Heun equation we make a partial fraction expansion of the coefficient functions and we obtain 
\begin{equation}\label{ODE1}
R^{''}_{\pm}(\rho)+p_{\pm}(\rho)R^{'}_{\pm}(\rho)+q_{\pm}(\rho)R_{\pm}(\rho)=0 
\end{equation}
with
\begin{equation}
p_{\pm}(\rho)=\frac{1/2}{\rho}+\frac{1/2}{\rho-1}-\frac{1}{\rho-c_\pm},\quad c_\pm=\mp i\frac{\lambda}{\mu}
\end{equation}
and
\begin{eqnarray}
q_{\pm}(\rho)&=&\Omega^2-\mu^2+\frac{\lambda^2}{\rho}+\frac{\alpha_{\pm}}{\rho-1}+\frac{\beta_{\pm}}{(\rho-1)^2}+\frac{\gamma_{\pm}}{\rho-c_\pm},\quad\alpha_{\pm}=2\Omega^2-\lambda^2-\mu^2-\gamma_{\pm},\\
\beta_{\pm}&=&\Omega^2\pm i\frac{\Omega}{2},\quad
\gamma_{\pm}=\mp\frac{\lambda\Omega}{i\lambda\pm\mu}.
\end{eqnarray}
By means of the transformation
\begin{equation}
R_{\pm}(\rho)=e^{-\sqrt{\mu^2-\Omega^2}\rho}\widetilde{R}_{\pm}(\rho)
\end{equation}
the differential equation (\ref{ODE1}) becomes
\begin{equation}\label{ODE2}
\widetilde{R}_{\pm}^{''}(\rho)+P_{\pm}(\rho)\widetilde{R}_{\pm}^{'}(\rho)+Q_{\pm}(\rho)\widetilde{R}_{\pm}(\rho)=0 
\end{equation}
with
\begin{equation}
P_{\pm}(\rho)=p_{\pm}(\rho)-2\sqrt{\mu^2-\Omega^2},\quad
Q_{\pm}(\rho)=\frac{\sigma}{\rho}+\frac{\widetilde{\alpha}_{\pm}}{\rho-1}+\frac{\beta_{\pm}}{(\rho-1)^2}+\frac{\widetilde{\gamma}_{\pm}}{\rho-c_\pm},
\end{equation}
where
\begin{equation}
\sigma=\lambda^2-\frac{\sqrt{\mu^2-\Omega^2}}{2},\quad
\widetilde{\alpha}_{\pm}=\alpha_{\pm}-\frac{\sqrt{\mu^2-\Omega^2}}{2},\quad 
\widetilde{\gamma}_{\pm}=\gamma_{\pm}+\sqrt{\mu^2-\Omega^2}.
\end{equation}
In order to eliminate terms like $(\rho-1)^{-2}$ in $Q_\pm$ we introduce the transformation
\begin{equation}
\widetilde{R}_{\pm}(\rho)=(\rho-1)^{\delta_\pm}\widehat{R}_{\pm}(\rho)
\end{equation}
with $\delta_\pm\in\mathbb{C}$ and hence, $\widehat{R}_{\pm}$ must satisfy the differential equation
\begin{equation}
\widehat{R}_{\pm}^{''}(\rho)+\mathfrak{p}_{\pm}(\rho)\widehat{R}_{\pm}^{'}(\rho)+\mathfrak{q}_{\pm}(\rho)\widehat{R}_{\pm}(\rho)=0 
\end{equation}
with
\begin{equation}
\mathfrak{p}_{\pm}(\rho)=\frac{1/2}{\rho}+\frac{2\delta_\pm+(1/2)}{\rho-1}-\frac{1}{\rho-c_\pm}-2\sqrt{\mu^2-\Omega^2},\quad
\mathfrak{q}_{\pm}(\rho)=\frac{\widehat{\sigma}_\pm}{\rho}+\frac{\widehat{\alpha}_{\pm}}{\rho-1}+\frac{\widehat{\beta}_{\pm}}{(\rho-1)^2}+\frac{\widehat{\gamma}_{\pm}}{\rho-c_\pm},
\end{equation}
where
\begin{equation}
\widehat{\sigma}_\pm=\sigma-\frac{\delta_\pm}{2},\quad
\widehat{\alpha}_{\pm}=\widetilde{\alpha}_{\pm}-2\delta_\pm\sqrt{\mu^2-\Omega^2}+\frac{\delta_\pm}{2}+\frac{\delta_\pm}{c_\pm-1},\quad
\widehat{\beta}_{\pm}=\delta_\pm^2-\frac{\delta_\pm}{2}+\beta_\pm,\quad
\widehat{\gamma}_{\pm}=\widetilde{\gamma}_{\pm}-\frac{\delta_\pm}{c_\pm-1}.
\end{equation}
Observe that $\widehat{\beta}_\pm=0$ whenever $\delta_\pm=\pm i\Omega$. Hence, by introducing the rescaled radial variable $\rho=r/(2M)$ and employing the ansatz
\begin{equation}\label{radspinor}
R_{\pm}(\rho)=(\rho-1)^{\pm i\Omega}e^{-\sqrt{\mu^2-\Omega^2}\rho}\widehat{R}_\pm(\rho),
\end{equation}
the radial system (\ref{r1}) and (\ref{r2}) decoupled into the following generalized Heun equations
\begin{equation}\label{GHE}
\widehat{R}_{\pm}^{''}(\rho)+\left(\sum_{n=0}^2\frac{1-\mu_{n,\pm}}{\rho-\rho_n}+\alpha\right)\widehat{R}_{\pm}^{'}(\rho)+\frac{\beta_{0,\pm}+\beta_{1,\pm}\rho+\beta_{2,\pm}\rho^2}{\prod_{n=0}^2(\rho-\rho_n)}\widehat{R}_{\pm}(\rho)=0
\end{equation}
where $\alpha=-2\sqrt{\mu^2-\Omega^2}$ and $\{0,\mu_{0,\pm}\}$, $\{0,\mu_{1,\pm}\}$, and $\{0,\mu_{2,\pm}\}$. Here, $\mu_{0,\pm}=1/2$, $\mu_{1,\pm}=(1/2)\mp 2i\Omega$, $\mu_{2,\pm}= 2$
 are the exponents associated to the simple singularities $\rho_0=0$, $\rho_1=1$, and $\rho_2=c_\pm$ with $c_\pm=\mp i\lambda/\mu$ while the point at infinity is an irregular singular point of rank at most one. Furthermore, we have
\begin{equation}
\beta_{0,\pm}=c_\pm \widehat{\sigma}_\pm,\quad
\beta_{1,\pm}=-c_\pm(\widehat{\alpha}_\pm+\widehat{\sigma}_\pm)-\widehat{\gamma}_\pm-\widehat{\sigma}_\pm,\quad
\beta_{2,\pm}=\widehat{\alpha}_\pm+\widehat{\gamma}_\pm+\widehat{\sigma}_\pm
\end{equation}
with
\begin{equation}
\widehat{\alpha}_\pm=2\Omega^2-\lambda^2-\mu^2-\frac{\sqrt{\mu^2-\Omega^2}}{2}\left(1\pm 4i\Omega\right)\mp i\frac{\Omega}{2},\quad
\widehat{\gamma}_\pm=\pm i\Omega+\sqrt{\mu^2-\Omega^2},\quad
\widehat{\sigma}_\pm=\lambda^2-\frac{\sqrt{\mu^2-\Omega^2}}{2}\mp i\frac{\Omega}{2}.
\end{equation}
Looking back at (\ref{radspinor}), we see that the problem of existence of bound states reduces to the question whether or not the generalized Heun equation (\ref{GHE}) admits polynomial solutions for $|\Omega|<\mu$, since then (\ref{radspinor}) would clearly satisfy the normalization condition  (\ref{seven}). To this purpose, let us first rewrite (\ref{GHE}) as follows
\begin{equation}
\rho(\rho-1)(\rho-c_\pm)\widehat{R}_{\pm}^{''}(\rho)+\mathfrak{P}_\pm(\rho)\widehat{R}_{\pm}^{'}(\rho)+\mathfrak{Q}_\pm(\rho)\widehat{R}_{\pm}(\rho)=0
\end{equation}
with $c_\pm$ as defined in the statement of the previous theorem and
\begin{eqnarray}
\mathfrak{P}_\pm(\rho)&=&(1-\mu_{0,\pm})(\rho-1)(\rho-c_\pm)+(1-\mu_{1,\pm})\rho(\rho-c_\pm)+(1-\mu_{2,\pm})\rho(\rho-1)+\alpha\rho(\rho-1)(\rho-c_\pm),\\
\mathfrak{Q}_\pm(\rho)&=&\beta_{0,\pm}+\beta_{1,\pm}\rho+\beta_{2,\pm}\rho^2.
\end{eqnarray}
Furthermore, suppose that
\begin{equation}
\widehat{R}_{\pm}(\rho)=\sum_{n=0}^\infty d_{\pm,n}(\rho-1)^n.
\end{equation}
In order that this infinite series stops at some fixed value of $n$, we first need to derive the recurrence relation satisfied by the coefficients $d_{\pm,n}$. To this purpose it results convenient to introduce the independent variable transformation $\tau=\rho-1$. Then, our differential equation becomes
\begin{equation}\label{equad}
\mathfrak{A}(\tau)\widehat{R}_{\pm}^{''}(\tau)+\mathfrak{P}_\pm(\tau)\widehat{R}_{\pm}^{'}(\tau)+\mathfrak{Q}_\pm(\tau)\widehat{R}_{\pm}(\tau)=0
\end{equation}
with
\begin{equation}
\mathfrak{A}(\tau)=\tau^3+a_{1,\pm}\tau^2+a_{2,\pm}\tau,\quad
\mathfrak{P}_\pm(\tau)=\alpha\tau^3+b_{1,\pm}\tau^2+b_{2,\pm}\tau+b_{3,\pm},\quad
\mathfrak{Q}_\pm(\tau)=\beta_{2,\pm}\tau^2+c_{1,\pm}\tau+c_{2,\pm},
\end{equation}
where
\begin{eqnarray}
a_{1,\pm}&=&2-c_\pm,\quad a_{2,\pm}=1-c_\pm,\\
b_{1,\pm}&=&\alpha(2-c_\pm)\pm 2i\Omega,\quad b_{2,\pm}=\alpha(1-c_\pm)-c_{\pm}(1\pm 2i\Omega)+\frac{1}{2}\pm 4i\Omega,\quad
b_{3,\pm}=(1-c_\pm)\left(\frac{1}{2}\pm 2i\Omega\right),\\
\beta_{2,\pm}&=&2\Omega^2-\mu^2\mp 2i\Omega\sqrt{\mu^2-\Omega^2},\quad
c_{1,\pm}=\beta_{1,\pm}+2\beta_{2,\pm},\quad
c_{2,\pm}=\beta_{0,\pm}+\beta_{1,\pm}+\beta_{2,\pm}.
\end{eqnarray}
Substituting $\widehat{R}_{\pm}(\tau)=\sum_{n=0}^\infty d_{\pm,n}\tau^n$ into (\ref{equad}) and shifting indices appropriately yield the following four term recurrence relation
\begin{equation}\label{*rr}
\varphi_1(n+1)d_{\pm,n+1}+\varphi_2(n)d_{\pm,n}+\varphi_3(n-1)d_{\pm,n-1}+\varphi_4(n-2)d_{\pm,n-2}=0,\quad n=0,1,2,\cdots
\end{equation}
with
\begin{eqnarray}
\varphi_1(\xi)&=& a_{2,\pm}\xi(\xi-1)+b_{3,\pm}\xi,\quad \varphi_2(\xi)=a_{1,\pm}\xi(\xi-1)+b_{2,\pm}\xi+c_{2,\pm},\\
\varphi_3(\xi)&=&\xi(\xi-1)+b_{1,\pm}\xi+c_{1,\pm},\quad\varphi_4(\xi)=\alpha\xi+\beta_{2,\pm}.
\end{eqnarray}
In order to find under which conditions we may have polynomial solutions of the form
\begin{equation}
\sum_{n=0}^N d_{\pm,n}\tau^n,\quad N=0,1,2,\cdots
\end{equation}
let $n=N+2$ in (\ref{*rr}). Then, we obtain
\begin{equation}
\varphi_1(N+3)d_{\pm,N+3}+\varphi_2(N+2)d_{\pm,N+2}+\varphi_3(N+1)d_{\pm,N+1}+\varphi_4(N)d_{\pm,N}=0,\quad N=0,1,2,\cdots
\end{equation}
For any fixed $N$ we will have a polynomial solution of degree $N$ whenever
\begin{equation}
\varphi_4(N)=0,\quad d_{\pm,N+1}=d_{\pm,N+2}=0.
\end{equation}
Bound states solutions will exist for those real values of $\Omega$ such that $|\Omega|<\mu$ and $\Omega$ satisfies simultaneously the above set of equations. It is not difficult to verify that the condition $\varphi_4(N)=0$ can be rewritten as $\alpha N+\beta_{2,\pm}=0$ and taking into account that $\alpha=-2\sqrt{\mu^2-\Omega^2}$ the same condition reads
\begin{equation}\label{ascia}
2\Omega^2-\mu^2-2N\sqrt{\mu^2-\Omega^2}=\pm 2i\Omega\sqrt{\mu^2-\Omega^2}.
\end{equation}
We argue now by contradiction. Let us suppose that for some $N\in\mathbb{N}\cup\{0\}$ there exists a bound state with energy $\Omega_N\in\mathbb{R}$ such that $|\Omega_N|<\mu$. This would imply that the l.h.s. of (\ref{ascia}) representing a real number should be equal to the r.h.s. of (\ref{ascia}) representing an imaginary number and hence, we have a contradiction. Therefore, there are no polynomial solutions for any $N=0,1,2,\cdots$.

\section{A deficiency index approach}
In this section we offer an alternative proof that no bound states exist for the Dirac equation in the Schwarzschild BH metric. This is achieved by constructing a suitable transformation of the radial system and showing that the deficiency indices of the transformed radial operator are zero. This is not surprising since the deficiency index of a differential operator counts the number of square integrable solutions. In order to be able to apply the Index Theorem in \cite{L} we bring the radial system (\ref{CRAS}) into the form of a symmetric Dirac system. This is done by transforming the radial spinors according to
\begin{equation}
(R_-(r),R_+(r))^T=(F(r)-iG(r),F(r)+iG(r))^T
\end{equation}
where $T$ denotes transposition and introducing a tortoise coordinate defined through
\begin{equation}
\frac{du}{dr}=\frac{r}{r-2M}
\end{equation}
whose solution is
\begin{equation}
u(r)=r+2M\ln{(r-2M)}.
\end{equation}
Note that $u\to+\infty$ as $r\to+\infty$ and $u\to-\infty$ as $r\to 2M^+$. If we set $\widehat{\Phi}=(F,G)^T$, the system (\ref{CRAS}) becomes
\begin{equation}\label{bbox}
(\mathcal{U}\widehat{\Phi})(u):=J\frac{d\widehat{\Phi}}{du}+B(u)\widehat{\Phi}=\omega\widehat{\Phi}
\end{equation}
with
\begin{equation}
J=\left(
\begin{array}{cc}
0 & 1\\
-1 & 0
\end{array}
\right),\quad
B(u)=\frac{\sqrt{\Delta_r(u)}}{r^2(u)}\left(
\begin{array}{cc}
m_e r(u) & \lambda\\
\lambda & -m_e r(u)
\end{array}
\right).
\end{equation}
According to the above transformations the integrability condition (\ref{seven}) for the radial spinors simplifies to
\begin{equation}\label{sstella}
(\widehat{\Phi},\widehat{\Phi})=\int_{-\infty}^{+\infty}du (F^2(u)+G^2(u))<+\infty.
\end{equation}
The formal differential operator $\mathcal{U}$ is formally symmetric because $J=-J^*$ and $B=B^*$. Let $\mathfrak{S}_{min}$ be the minimal operator associated to $\mathcal{U}$ such that $\mathfrak{S}_{min}$ acts on the Hilbert space $L^2(\mathbb{R},du)$ with respect to the scalar product $(\cdot,\cdot)$ introduced in (\ref{sstella}). Then, the operator $\mathfrak{S}_{min}$ with domain of definition $D(\mathfrak{S}_{min})=C_0^\infty(\mathbb{R})^2$ such that $\mathfrak{S}_{min}\widehat{\Phi}:=\mathcal{U}\widehat{\Phi}$ for $\widehat{\Phi}\in D(\mathfrak{S}_{min})$ is densely defined and closable. Let $\mathfrak{S}$ denote the closure of $\mathfrak{S}_{min}$. In order to apply Neumark's decomposition method \cite{Neu}we introduce minimal operators $\mathfrak{S}_{min,\pm}$ associated to $\mathcal{U}$ when the latter is restricted to the half-lines $I_+=[0,+\infty)$ and $I_-=(-\infty,0]$, respectively. Moreover, we consider $\mathfrak{S}_{min,\pm}$ acting on the Hilbert space $L^2(I_\pm,du)$ with respect to the scalar product $(\cdot,\cdot)$. The operators $\mathfrak{S}_{min,\pm}$ with domains of definition $D(\mathfrak{S}_{min,\pm})=C_0^\infty(I_\pm)^2$ and such that $\mathfrak{S}_{min,\pm}\widehat{\Phi}_\pm:=\mathcal{U}\widehat{\Phi}_\pm$ for $\widehat{\Phi}_\pm\in D(\mathfrak{S}_{min,\pm})$ are again densely defined and closable. Furthermore, $\mathcal{U}$ is in the limit point case (l.p.c.) at $\pm\infty$. This can be seen as follows. First of all, we recall that since the limit point and limit circle cases are mutually exclusive, we can determine the appropriate case if we examine the solution of (\ref{bbox}) for a single value of $\omega$. Hence, without loss of generality we set $\omega=0$ and consider the system
\begin{equation}\label{aascia}
J\frac{d\widehat{\Psi}}{du}+B(u)\widehat{\Psi}=0.
\end{equation} 
For $u\to-\infty$ it is straightforward to verify that the radial variable $r$ admits the asymptotic expansion
\begin{equation}
r=2M+e^{\frac{u}{2M}}+\mathcal{O}\left(e^{\frac{u}{M}}\right).
\end{equation}
Furthermore, in the same asymptotic limit the matrix $B(u)$ has the following decomposition
\begin{equation}
B(u)=B_0(u)+B_1(u),\quad
B_0(u)=e^{\frac{u}{4M}}\left(
\begin{array}{cc}
m_e & \frac{\lambda}{2M}\\
\frac{\lambda}{2M} & -m_e
\end{array}
\right),\quad |B_1(u)|\leq C e^{\frac{3u}{4M}},
\end{equation}
where $C$ is a positive constant. Finally, it can be easily checked that the system (\ref{aascia}) has asymptotic solution for $u\to-\infty$ given by
\begin{equation}\label{fr1}
F(u)=\mbox{exp}\left(2\gamma e^{\frac{u}{2M}}\right)\left(
\begin{array}{c}
1\\
0
\end{array}
\right)+\mathcal{O}\left(e^{\frac{3u}{4M}}\right),\quad
G(u)=\mbox{exp}\left(-2\gamma e^{\frac{u}{2M}}\right)\left(
\begin{array}{c}
0\\
1
\end{array}
\right)+\mathcal{O}\left(e^{\frac{3u}{4M}}\right),\quad\gamma=\sqrt{\lambda^2+4M^2 m_e^2}.
\end{equation}
When $u\to+\infty$, the tortoise coordinate becomes
\begin{equation}
u=r+2M\ln{r}+\mathcal{O}\left(\frac{1}{u}\right)
\end{equation}
from which we find that $r$ can be expressed as a function of $u$ as follows
\begin{equation}
r(u)=\mbox{exp}\left(-\left[W\left(\frac{e^{\frac{u}{2M}}}{2M}\right)-\frac{u}{2M}\right]\right)+\mathcal{O}\left(\frac{1}{u}\right).
\end{equation} 
Here, $W$ denotes the so-called Lambert function \cite{LAM}. Finally, expanding $W$ asymptotically for $u\to +\infty$ as in \cite{Dub} yields
\begin{equation}
r=u-2M\ln{u}+\mathcal{O}\left(\frac{1}{u}\right).
\end{equation}
When $u\to+\infty$ the matrix $B(u)$ admits the decomposition
\begin{equation}
B(u)=\widetilde{B}_0(u)+\widetilde{B}_1(u),\quad
\widetilde{B}_0(u)=\left(
\begin{array}{cc}
m_e & 0\\
0 & -m_e
\end{array}
\right),\quad |\widetilde{B}_1(u)|\leq \frac{D}{u}
\end{equation}
for some positive constant $D$. Hence, the system (\ref{aascia}) admits the following solution for $u\to+\infty$
\begin{equation}\label{fr2}
F(u)=e^{m_e u}\left(
\begin{array}{c}
1\\
0
\end{array}
\right)+\mathcal{O}\left(\frac{1}{u}\right),\quad
G(u)=e^{-m_e u}\left(
\begin{array}{c}
0\\
1
\end{array}
\right)+\mathcal{O}\left(\frac{1}{u}\right).
\end{equation}
By inspecting (\ref{fr1}) and (\ref{fr2}) we can immediately conclude that the differential operator $\mathcal{U}$ is in the l.p.c. at $\pm\infty$. Hence, the operators $\mathfrak{S}_{min,\pm}$ are essentially self-adjoint. Let $\mathfrak{S}_\pm$ denote the closure and $N(\mathfrak{S}_\pm)$ the corresponding deficiency indices. If $\nu_\pm$ denotes the number of positive and negative eigenvalues of the matrix $iJ$, then, since $\nu_+=1=\nu_-$, Theorem~5.2 in \cite{L} implies that $N_\pm(\mathfrak{S}_+)=1=N_\pm(\mathfrak{S}_-)$. Since zero is the only solution of (\ref{aascia}) in $L^2(\mathbb{R},du)$ the original system (\ref{bbox}) is definite on $I_+$ and $I_{-}$ in the sense of \cite{L}. Finally, (5.11a) in Proposition~5.4 in \cite{L} yields that the deficiency indices for the operator $\mathfrak{S}$ are
\begin{equation}
N_\pm(\mathfrak{S})=N_\pm(\mathfrak{S}_+)+N_\pm(\mathfrak{S}_-)-2=0.
\end{equation}
This implies that the radial system (\ref{bbox}) does not posses any square integrable solution on the whole real line and therefore no bound states for the Dirac equation in the Schwarzschild BH metric are allowed.

\section{Conclusions}
In this paper we have focused on the existence of fermionic bound states around a black hole. 
Although such proofs are not new, our methods in Section III and IV are original and suitable to show the absence of the bound states 
in particular for the case of $\omega < m_e$ and for any $\omega$ in the case the index theorem is used.. Claims regarding the existence
of such bound states have been made in the literature, but were based on approximation methods.
We used two different exact methods to complete the proof that regardless the value of the particle's mass the fermionic bound states do not exist. 
This means also that the aforementioned approximations were not adequate to study bound states in a curved background.
For those who arrive at the same conclusion as us (i.e. absence of bound states in the black hole metric) but use the full atlas (including
the inside of the black hole) the physical explanation for the absence of bound states is that the central singularity acts as a current sink \cite{Lasen}.
Since we are working in the Schwarzschild coordinates we think that the presence of the horizon is sufficient to explain the phenomenon. 
To appreciate the physical relevance of the result let us highlight the fact that the absence of bound states around a black hole is a proper quantum effect in the sense 
explained hereafter. 
In classical electrodynamics a charged particle in motion around a point-like object should emit radiation, making atoms unstable. 
The resolution of this problem takes place in Quantum Mechanics which predicts the existence of stable, non-radiating bound states. 
The opposite seems to happen in gravitation. Classically, a particle can orbit around a black hole along a geodesic outside the horizon and such an orbit is stable 
\cite{Ch}. 
But relativistic wave mechanics and spin (i.e. Dirac equation in curved spacetime) breaks down completely the classical picture and as 
a consequence, no stable orbits exist. We can say that this happens due to the finite probability finding the particle inside the horizon. 
It is quite natural to ask whether this result continues to hold in the case the gravity is that of the Fermion itself or in the presence of 
other effects such as those due to torsion. 
In general, we could raise the question if the results persist in modified versions of gravity.
These are questions we will study elsewhere. 

Let us shed some light on the above mentioned quantum effect drawing some analogies from quantum mechanics. Consider a classical hydrogen atom.
By Gauss law the electron orbiting outside the proton radius will not ``know'' whether the proton has a structure (finite size) or if it is a point-like object.
The electron experiences only the Coulomb force. This picture changes in quantum mechanics. The wave function is defined here over the whole space
and as a consequence we have to specify the full interaction potential. The choice of the Coulomb potential means that we opt for a point-like proton
and the choice to include the finite size of the proton will lead to a different result (see \cite{Flugge} for a simplified version of this picture and
\cite{Breit} for a more sophisticated approach). We encounter a similar situation with an electron around a black hole. Classically, the electron
cannot ``know'' whether it moves (in stable orbits) around a black hole or just in a spherically symmetric metric of a star. In quantum mechanics,
just in analogy to the hydrogen atom where the electron experiences also the ``inside-potential''  of the proton, the electron described 
by a Dirac equation coupled to a black hole metric will ``know'' about the existence of a horizon. The latter fact is the reason why we encounter
no bound states. Whereas in the hydrogen atom this effect is microscopic, in the black hole metric it is a macroscopic quantum effect which by itself
is interesting.
Both effects as seen from a classical perspective have to do also with non-locality which is best exemplified in the black hole case.
We could start putting the electron at a distance from the horizon at which classically we would expect a stable orbit.
What we should see due to the fact that a stable orbit essentially does not exist is the decay of the same. The non-locality 
shows up since the electron at a classical distance will ``know'' if there is a horizon or not.

\section*{Acknowledgments}
We would like to thank the anonymous referees for their useful comments and suggestions.


\end{document}